\documentclass[a4paper,
              ]{jacow}

\makeatletter%                           % test for XeTeX where the sequence is by default eps-> pdf, jpg, png, pdf, ...
\ifboolexpr{bool{xetex}}                 % and the JACoW template provides JACpic2v3.eps and JACpic2v3.jpg which might generates errors
 {\renewcommand{\Gin@extensions}{.pdf,%
                    .png,.jpg,.bmp,.pict,.tif,.psd,.mac,.sga,.tga,.gif,%
                    .eps,.ps,%
                    }}{}
\makeatother

\ifboolexpr{bool{xetex} or bool{luatex}} % test for XeTeX/LuaTeX
 {}                                      % input encoding is utf8 by default
 {\usepackage[utf8]{inputenc}}           % switch to utf8

\usepackage[USenglish]{babel}

\ifboolexpr{bool{jacowbiblatex}}%        % if BibLaTeX is used
 {%
  \addbibresource{jacow-test.bib}
  \addbibresource{biblatex-examples.bib}
 }{}

\usepackage{epstopdf}

\begin{document}

\title{Comparison of Coherent Smith-Purcell radiation and Coherent Transition Radiation\thanks{The authors are grateful for the funding received from the French ANR (contract ANR-12-JS05-0003-01) and the IDEATE International Associated Laboratory (LIA) between France and Ukraine.}}

\author{Vitalii Khodnevych\textsuperscript{1}\thanks{hodnevuc@lal.in2p3.fr}, Nicolas Delerue~\thanks{delerue@lal.in2p3.fr} 
\\LAL, Univ. Paris-Sud, CNRS/IN2P3, Universit\'e Paris-Saclay, Orsay, France.
\\ Oleg Bezshyyko, Taras Shevchenko National University of Kyiv, Kyiv
\\ \textsuperscript{1}also at National Taras Shevchenko University of Kyiv, Kyiv, Ukraine
  }

\maketitle

\begin{abstract}
Smith-Purcell radiation and Transition Radiation are two radiative phenomenon that occur in charged particles accelerators. For both the emission can be significantly enhanced with sufficiently short pulses and both can be used to measure the form factor of the pulse. We compare the yield of these phenomenon in different configurations and look at their application as bunch length monitors, including background filtering and rejection. We apply these calculations to the specific case of the CLIO Free Electron laser
\end{abstract}

\section{Introduction}
\subsection{Coherent Transition Radiation (CTR)}
When a relativistic charged particle crosses the interface between two media of different dielectric properties, transition radiation (TR) is emitted. This process was calculated analytically by Ginzburg and Frank \cite{GF}.
\begin{equation}
\frac{d^2I_{GF}}{d\omega d\Theta}=\frac{q_0^2}{4\pi^3\epsilon_0c}\frac{\beta^2sin^2\Theta}{(1-\beta^2cos^2\Theta)^2},
\label{eq:eq2}
\end{equation}
where $q_0$ is electron charge, $\epsilon_0$ -- vacuum permittivity, c is the speed of light, $\beta$  is relativistic velocity and $\Theta$ is
the observation angle.

 We use their formula with virtual-quanta method to compute the backward TR from a finite screen\cite{GFa} 
 \begin{equation}
\frac{d^2I_{disk}}{d\omega d\Theta}=\frac{d^2I_{GF}}{d\omega d\Theta}[1-T(\gamma,\omega a,\Theta)]^2
\label{eq:eq3}
\end{equation}
where T is correction tern to finite size of the screen. This gives the single electron yield (SEY). From the SEY the  whole spectrum can be derived using the following formula:
\begin{equation}
\frac{d^2I}{d\omega d\Theta}=\frac{d^2I_1}{d\omega d\Theta}[N+N(N-1)F(\omega)]
\label{eq:eq1}
\end{equation}
Where N is the number of electrons in the bunch and $F(\omega)$ is the form factor of the time profile of the bunch. Using phase recovery methods, such as Kramers-Kronig or Hilbert~\cite{IPAC14}, it is possible to recover  the phase and then the time profile of the bunch.\par

\subsection{Coherent Smith-Purcell Radiation (CSPR)}
The same formula, but with a different SEY is used for calculation of the Smith-Purcell radiation (SP) spectrum. SP radiation occurs when a charged particle move above a metallic periodic structure. Unlike TR, SP has the advantage  that the emitted radiation is not concentrated in a small observation angle ($\Theta_{max} 	\simeq1/\gamma$), but spread in angle. The wavelength of the radiation for SP depends on the observing angle according to the following:
\begin{equation}
\lambda=\frac{l}{n}(\frac{1}{\beta}-cos\Theta)
\label{eq:lmab}
\end{equation}
where l is the grating period, n is the order of radiation, $\Theta$ is the observation angle and $\beta$ is the relativistic velocity.\par
To calculate the SEY and the total spectrum for SP effect, the gfw code was used~\cite{GD}. The calculation is based on the surface current model. Taking into account the fact that the grating have a finite width, the energy per solid angle for a single electron can be written as:
\begin{equation}
\frac{dI}{d\Omega}=2\pi e^2\frac{Z}{l^2}\frac{n^2\beta^3}{(1-\beta cos\Theta)^3}R^2
\end{equation}
where Z is the grating length, e is the electron charge and $R^2$ is the grating efficiency factor.\par

\section{Comparison}
\subsection{Single Electron Yield}

\begin{figure}[htb]
  \centering
  \includegraphics[width=0.9\linewidth]{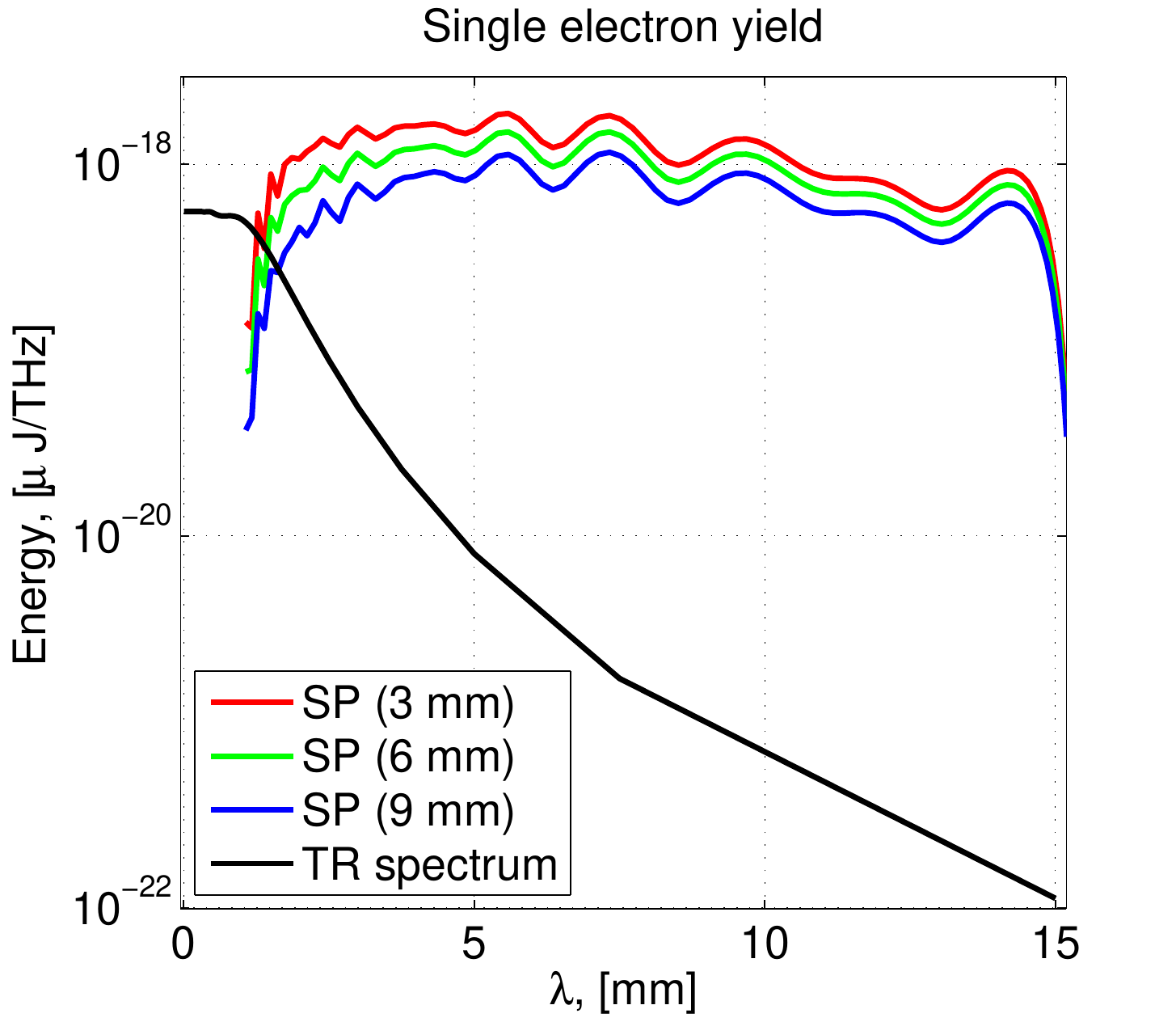}
  \caption{Single electron yield for TR and SP. The screen diameter for TR is 40mm. SP SEY is presented for different beam-grating separation (\SIlist{3;6;9}{mm}). The grating used here is   $40\times180$ \si{mm^2} with 8 mm pitch and $30^o$ blaze angle.  }
  \label{sey}
\end{figure}

Calculation of the SEY for SP and TR are presented on figure~\ref{sey}. The screen used in this case for TR is a metalic disk with radius 20 mm and the grating for SP has a length of 180 mm and a width of 40 mm  with a pitch of 8 mm  and a 30 degrees blaze angle. The beam-grating distance (from center of  the beam to the top of the teeth) is 3 mm. The lorentz factor $\gamma$ for both cases is 118 (60.3~MeV). The other parameters were taken from the CLIO accelerator simulation in ASTRA reported in~\cite{clio}. \par

\begin{figure}[htbp]
  \centering
  \includegraphics[width=0.9\linewidth]{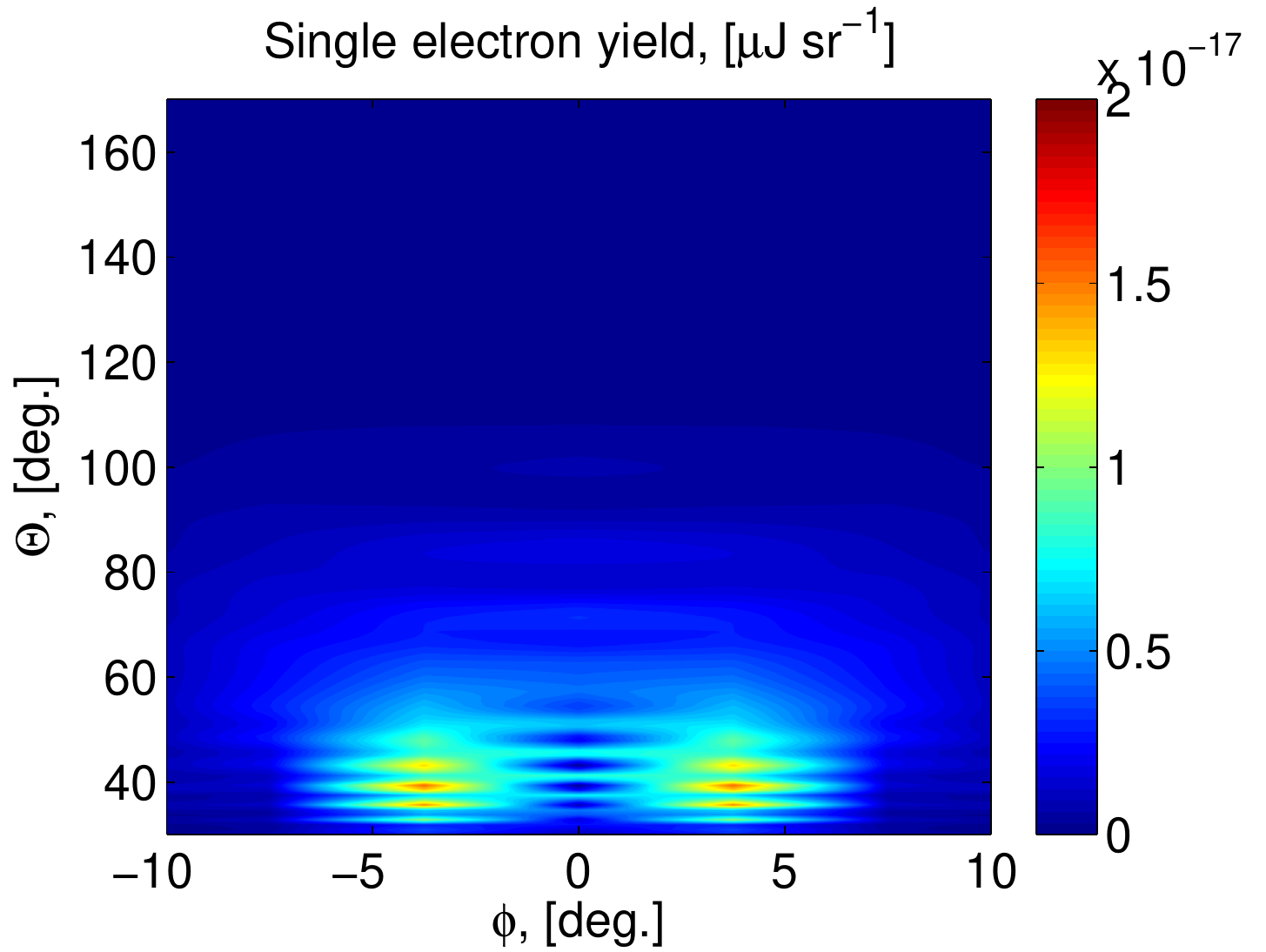}
  \caption{SEY for SP effect. Grating  $40\times180$ \si{mm^2}    with 8~mm pitch and $30^o$ blaze angle.  }
  \label{sey2d}
\end{figure}

The SEY spatial energy distribution for SP and TR are also significantly different. The spatial distribution of the energy per solid angle and per grating length  is presented on figure~\ref{sey2d}.
\begin{figure}[!htb]
  \centering
  \includegraphics[width=0.9\linewidth]{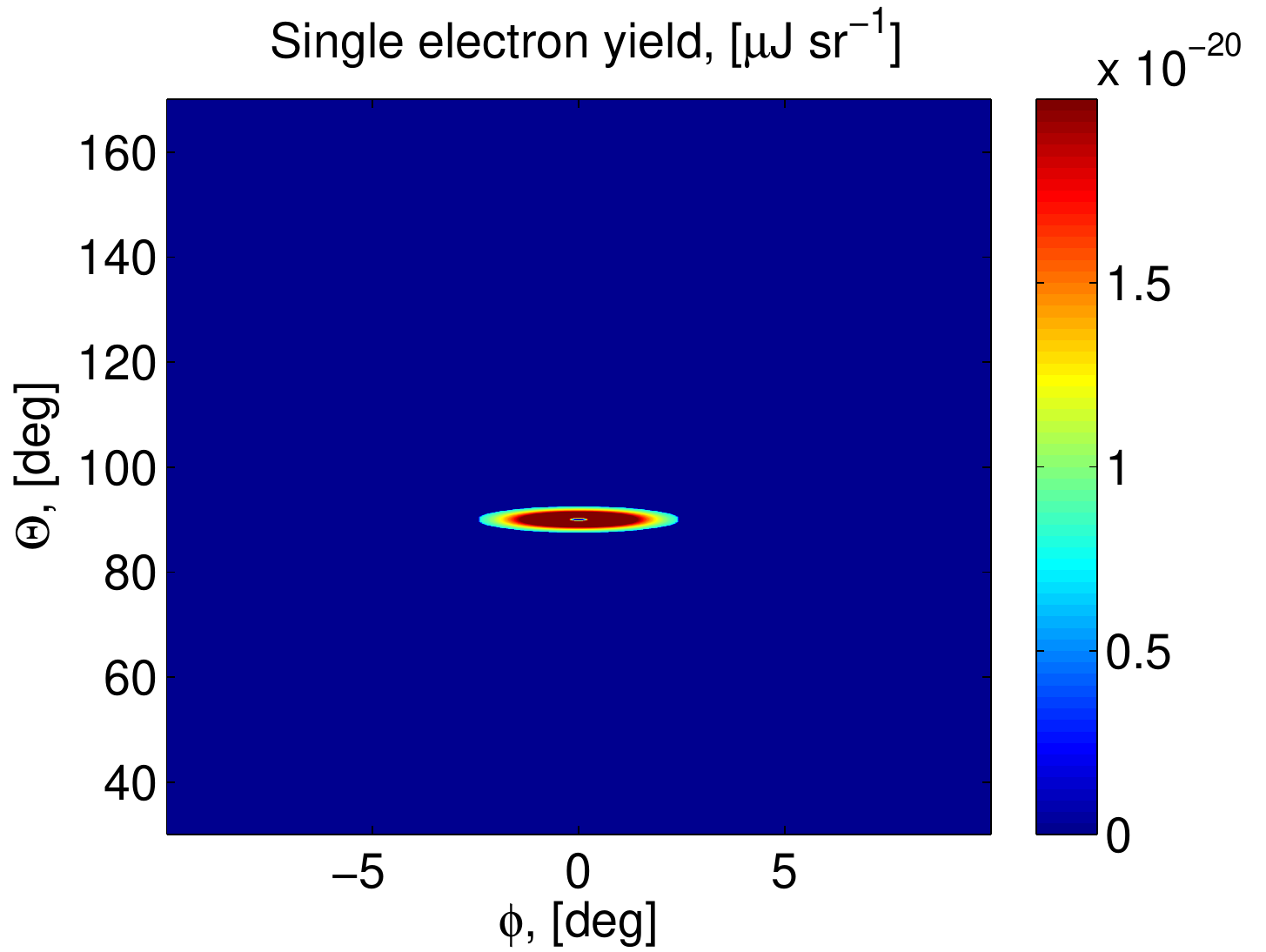}
  \caption{SEY for TR effect. Screen is turned at $45^o$ to beam propagation direction and have diameter 40 mm.}
  \label{sey2dctr}
\end{figure}
\subsection{Coherent Radiation}
%So, as both CTR and Smith-Purcell radiation depends from form factor of the time profile of the bunch (see eq. \ref{eq:eq1}),so they both  can be used to diagnose longitudinal beam profile. 
%Using the profile shape information from~\cite{clio} and the SEY, we can predict the spatial distribution of the energy for both effects. 
The calculation of coherent radiation was done with the same parameters than for the SEY. For others grating this distribution would be different, but this gives us an approximate space distribution of the CSPR. From these simulations, we can conclude  that most of the radiation is confined in  approximatively $\pm$\ang{6} in azimuthal ($\phi$)  angle. So a standard 50~mm parabolic mirror at a distance of 300~mm  from the grating will collect most of the radiation.\par
To choose the most appropriate grating pitch, one should use the condition given in equation~\ref{eq:lmab}.  For maximum emission at 90 deg. the formula \label{eq:pitch_pulselength} is applicable.
\begin{equation}
l=\frac{2\pi c}{2\sqrt{2ln(2)}}p_{t} 	\approx 8\times10^{8}p_{t}
\label{eq:pitch_pulselength}
\end{equation}
where l is grating pitch in meters and $p_{t}$ is the bunch FWHM in seconds.

\begin{figure}[!htbp]
  \centering
  \includegraphics[width=0.9\linewidth]{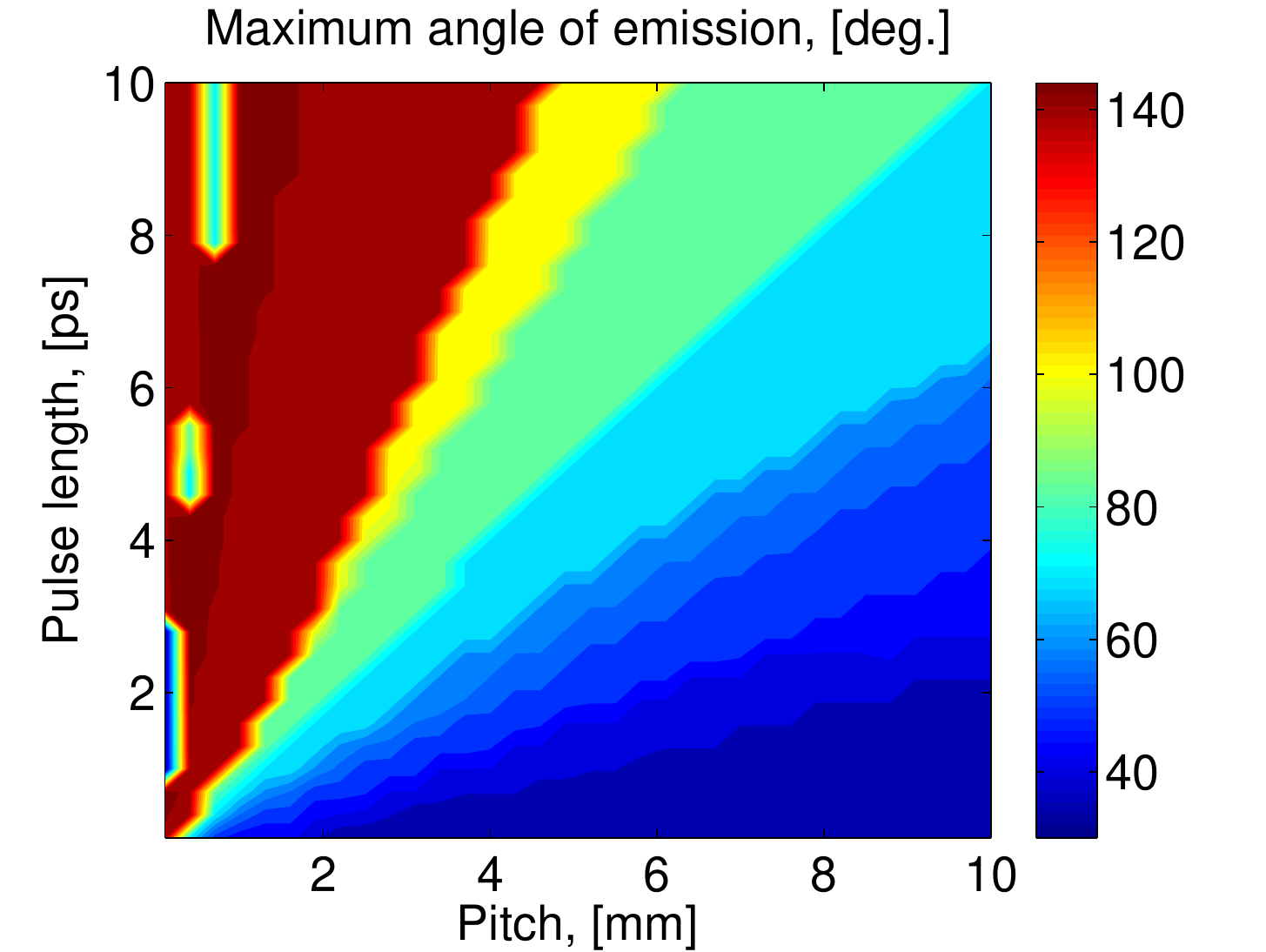}
  \caption{Maximum angle of emission for SP effect as function of pulsewidth and grating pitch.  Grating  $40\times180$ \si{mm^2} with  $30^o$ blaze angle. The beam-grating separation is 3~mm.}
  \label{mae}
\end{figure}

\begin{figure}[!htbp]
  \centering
  \includegraphics[width=0.9\linewidth]{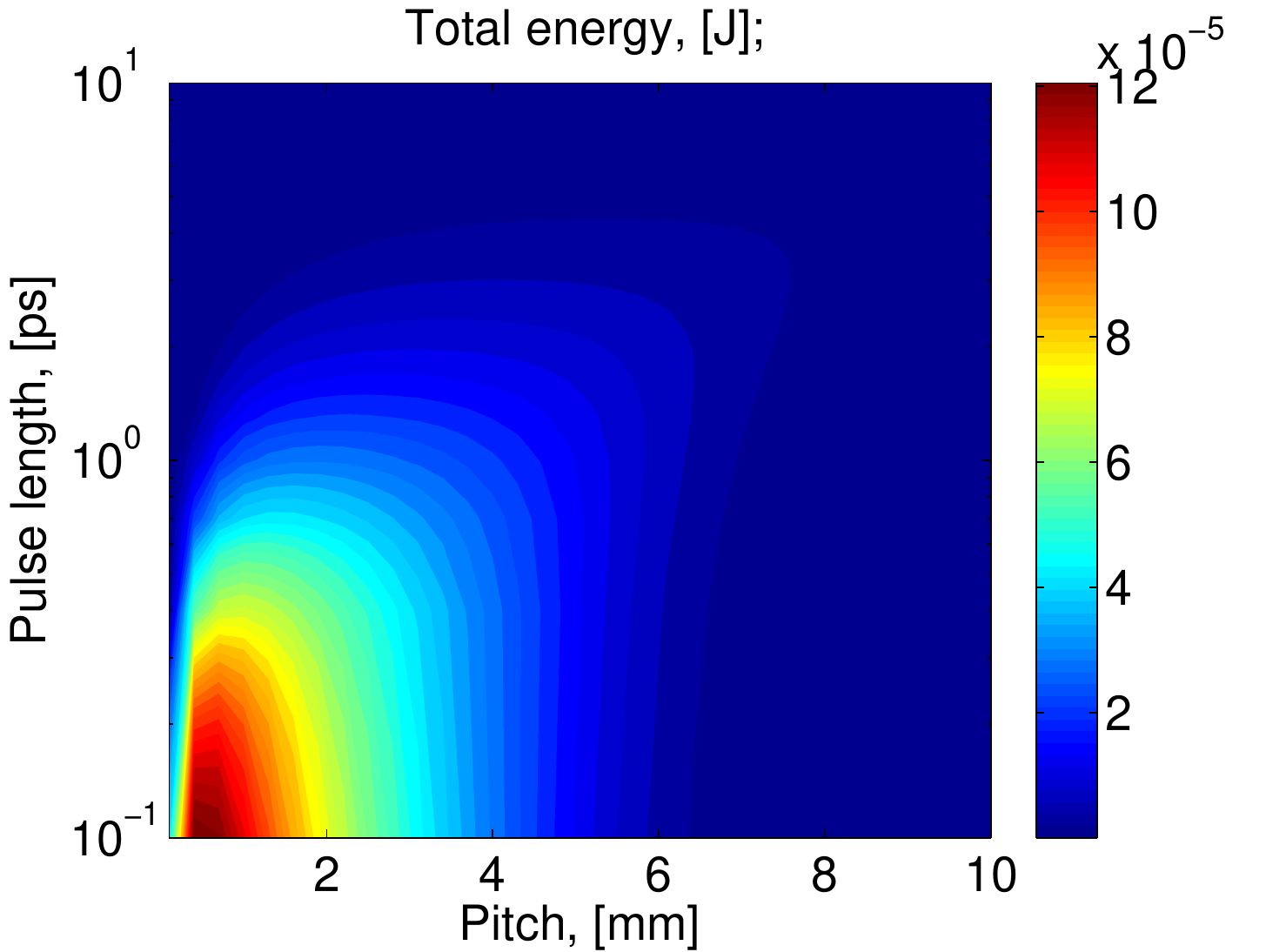}
  \caption{Total energy for SP effect presented as function of pulsewidth and grating pitch. Grating is  $40\times180$ \si{mm^2} with  $30^o$ blaze angle. }
  \label{Epp}
\end{figure}

The relation between the the grating pitch, the pulse length and the angle of maximum emission is given on figure~\ref{mae}. We can see that the green band ($90^\circ$ emission) follows the rule given in equation~\ref{eq:pitch_pulselength}. We can also look at the total energy emitted by the grating as a function of the pulse length and the grating energy. This is shown on figure~\ref{Epp}. On this figure the transition between the coherent and incoherent regime can clearly be seen.

%For same parameters set we also calculate wavelength that correspond to maximum of emission (see fig. \ref{wme}). This dependence is completely in agreemets with coherence condition \ref{eq:lmab}.\par
%\begin{figure}[!htb]
%  \centering
%  \includegraphics[width=0.9\linewidth]{plots/WME.eps}
%  \caption{Wavelength that correspond to maximum of emission. Grating $40\times180$mm with  $30^o$ blaze angle.}
%  \label{wme}
%\end{figure}

%incoherence

\begin{figure}[!tbp]
  \centering
  \includegraphics[width=0.9\linewidth]{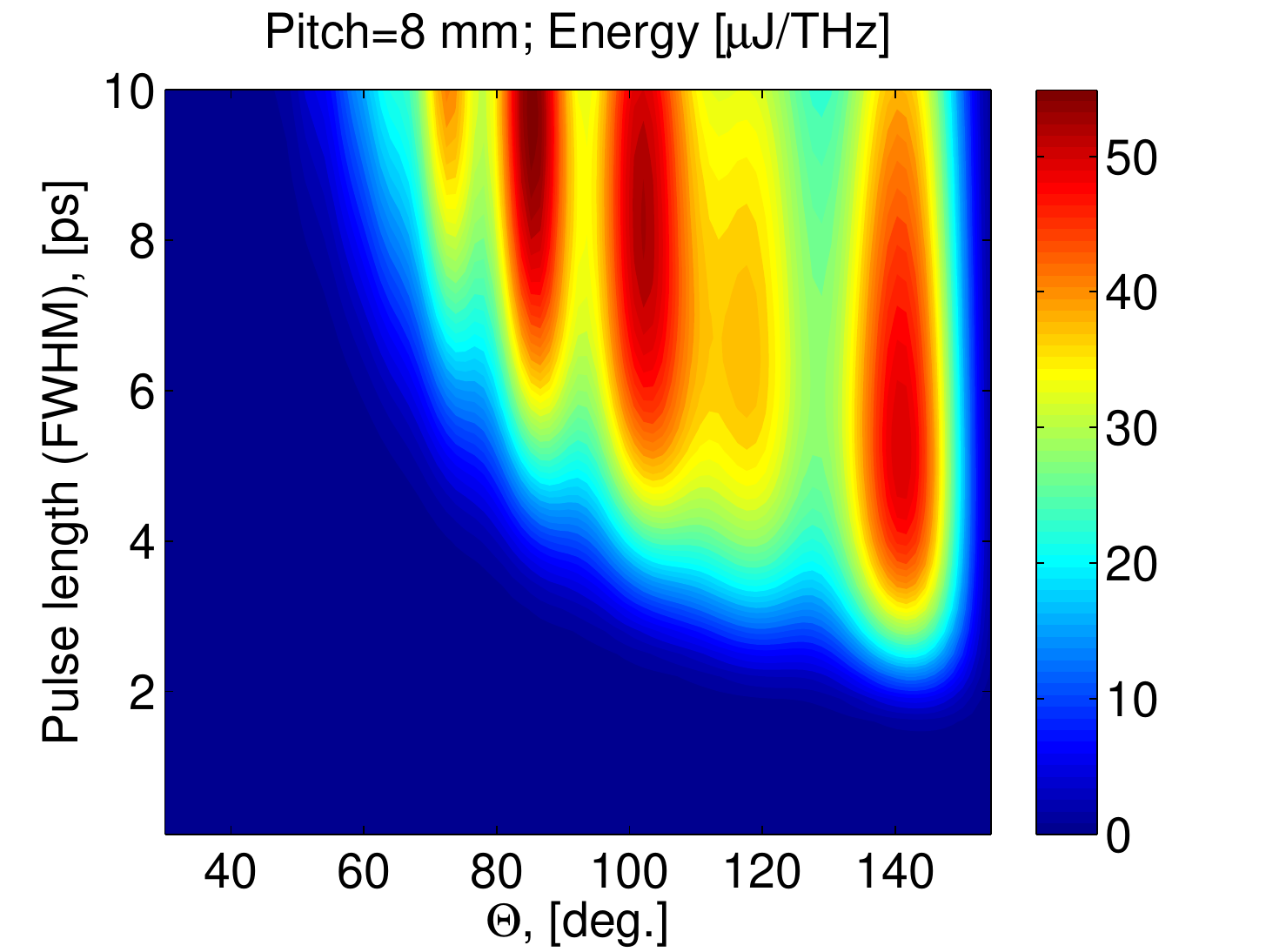}
  \caption{Evolution of  the CSPR spectrum with a constant grating pitch and a wavelength changing with the angle. The grating dimensions are $40\times180$~\si{mm^2} with \SI{8}{mm} pitch and \ang{30} blaze angle. }
  \label{f14}
\end{figure}

\begin{figure}[!tbp]
  \centering
  \includegraphics[width=0.9\linewidth]{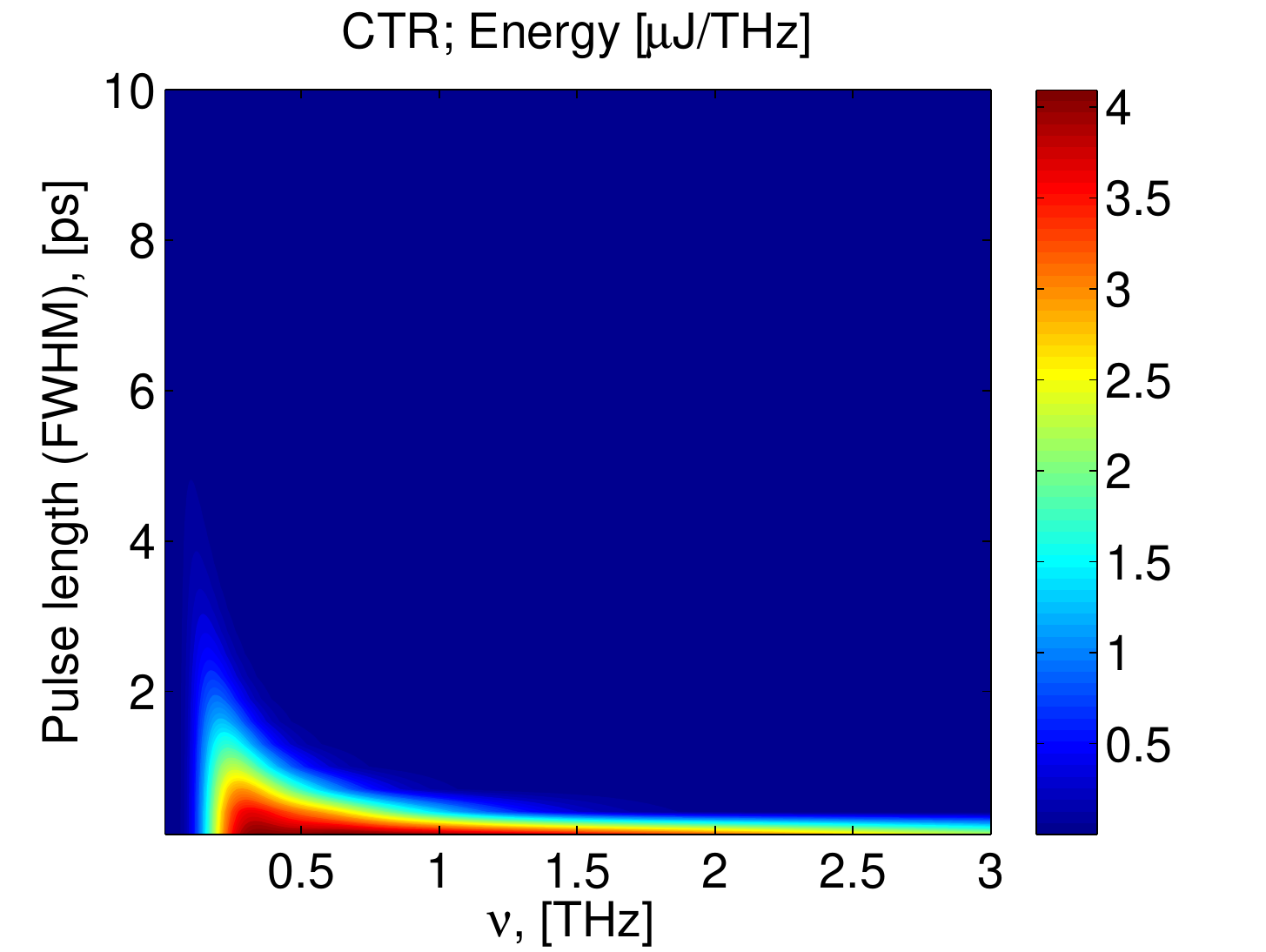}
  \caption{Evolution of CTR spectrum with  changing wavelength. The screen has a diameter of \SI{40}{mm}.}
  \label{f15}
\end{figure}

In the case of CSPR as the pulse length change the angular distribution of the energy will also change. This is shown on figure~\ref{f14}. Same dependence for CTR is shown on  figure~\ref{f15}.

%Evolution of  spectrum for cetain pulselength with pitch is presented on figure \ref{f13}. By changing pitch, we change sampling point in form factor and live region of interest (sampled amplitudes are almost zero) and in this case incoherent part start to dominate. Evolution of  spectrum for cetain grating pitch with pulselength is presented on figure \ref{f14}. This situation is similar to previous, except we change in formula \ref{eq:lmab} $\lambda$ instead grating pitch. So in this case for certain sampling points we change width of form factor.\par
%\begin{figure}[!htb]
%  \centering
%  \includegraphics[width=0.9\linewidth]{plots/constplen.eps}
 % \caption{Evolution of SP spectrum with constant pulselength and changing grating pitch. Grating $40\times180$mm with  $30^o$ blaze angle.}
%  \label{f13}
%\end{figure}

\begin{figure}[!bp]
  \centering
  \vspace{1.8cm}
  \includegraphics[width=0.8\linewidth]{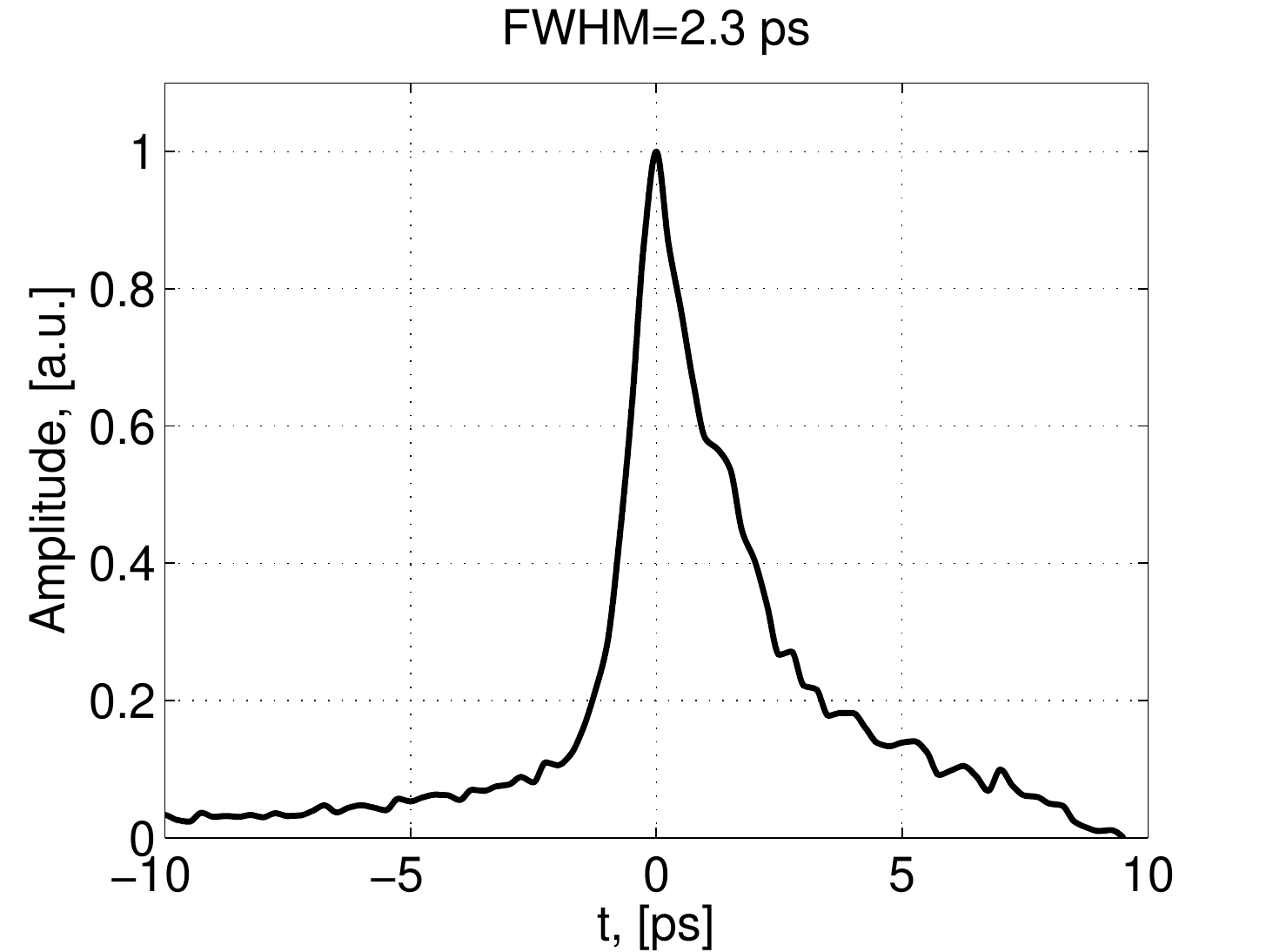}
  \caption{Profile of the bunch at the exit of the CLIO accelerating section (see~\cite{clio}).}
  \label{Prof1}
\end{figure}

\begin{figure}[!tbp]
  \centering
  \includegraphics[width=0.8\linewidth]{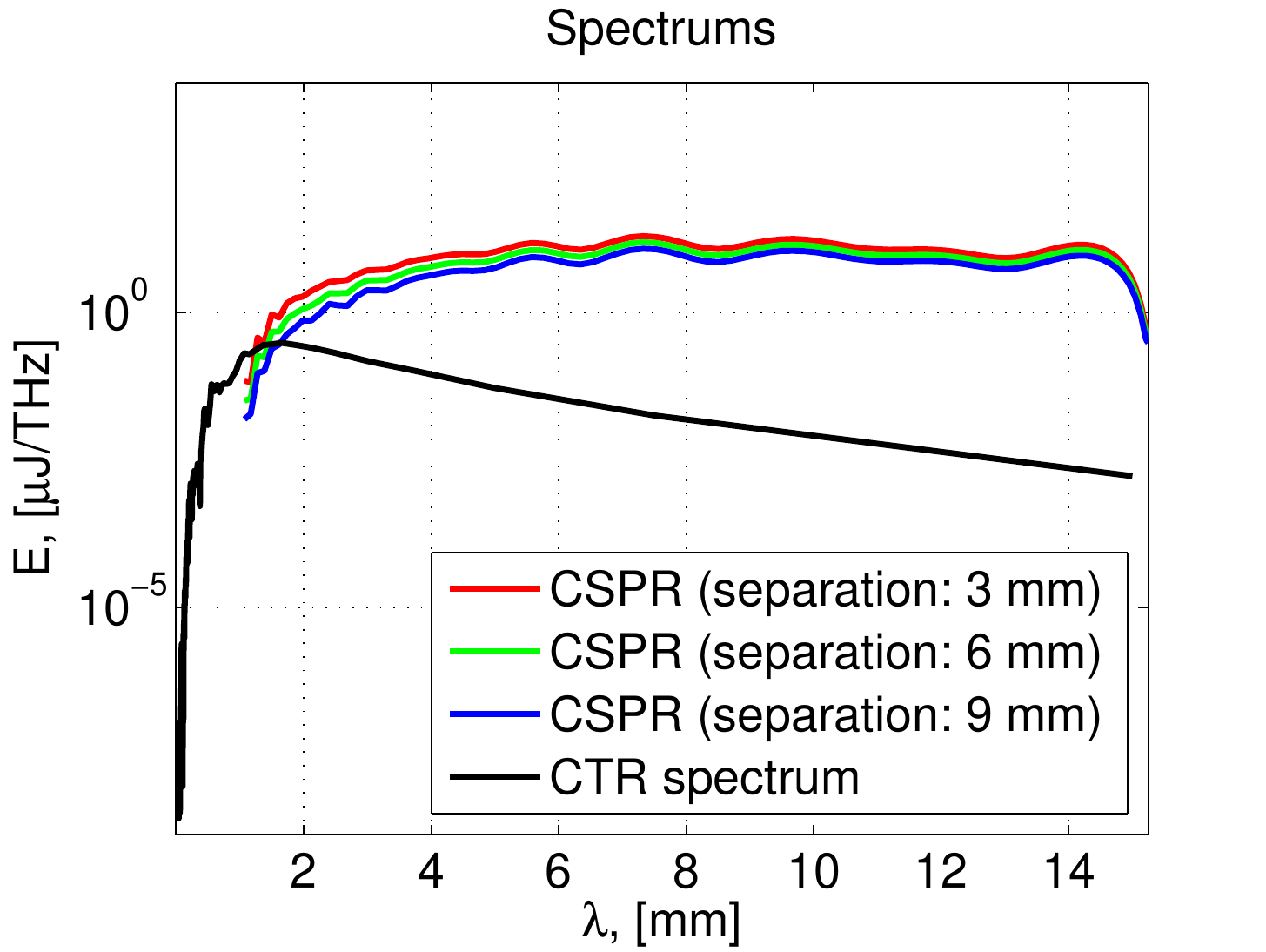}
  \caption{CSPR and CTR energy density as function of wavelength. The CSPR spectrum is presented for  beam-grating separations of \SIlist{3;6;9}{mm}. The grating dimensions are $40\times180$ \si{mm^2} with \SI{8}{mm} pitch and \ang{30} blaze angle. The screen diameter for TR is \SI{40}{mm}. The signal is measured as integrated with a \SI{50}{mm} diameter parabolic mirror located \SI{300}{mm} from the beam axis.}
  \label{spctr}
\end{figure}

Using the bunch profile predicted for the CLIO Free Electron Laser~\cite{clio}, as shown on figure~\ref{Prof1} we can predict the spectrum for both CSPR and CTR as shown on figure~\ref{spctr}. We can see that the intensity of the CTR signal is  lower, but it is concentrated in a small solid angle. For CSPR the signal intensity depends on the beam-grating separation.

\section{Conclusion}
We have studied both CSPR and CTR and studied how to optimize the experimental parameters. Using the CLIO parameters we expect a signal (in the range 0.03-3 THz [ 0.1 - 10 mm]) of \SI{8.37e-7}{J} for CSPR and \SI{7.35e-08}{J} for CTR.\par

%\begin{figure}[!htb]
 % \centering
%  \includegraphics[width=0.9\linewidth]{plots/TR.eps}
%  \caption{TR distribution in angle and wavelength space. Screen diameter for TR is 40mm. }
%  \label{tr}
%\end{figure}%

%\columnbreak

%\clearfigure

%\begin{figure}[!htb]
% \centering
 % \includegraphics*[width=70mm]{THPME088f3.eps} \\
%  \includegraphics*[width=70mm]{THPME088f4.eps} 
%  \caption{$\Delta_{FWXM}$  (top) and $\chi^2$ (bottom) distribution of our 1000 simulations reconstructed using the Hilbert transform method.}
%   \label{profiles_stats_hilbert}
%\end{figure}

%\bibliographystyle{unsrt}
%\bibliography{biblio}

%%\begin{thebibliography}{9}   % Use for  1-9  references
%\begin{thebibliography}{99} % Use for 10-99 references

%\bibitem{accelconf-ref}
%	C. Petit-Jean-Genaz and J. Poole,
%	``JACoW, A service to the Accelerator Community,''
%	EPAC'04, Lucerne, July 2004, THZCH03,  p.~249,
%	\url{http://www.JACoW.org/e04/papers/THZCH03.PDF}

% \end{thebibliography}

\end{document}